\documentclass[useAMS,usenatbib]{mn2e}
\usepackage{graphicx}
\def\rmscr#1{{\hbox{\rm \scriptsize #1}}}
\def\rmmat#1{{\hbox{\rm #1}}}
\def\d{\rmmat{d}}

\begin{document}
\title[White-Dwarf Kicks IV]{Constraining white-dwarf kicks in globular
  clusters : IV.~Retarding Core Collapse}
\date{Accepted 2009 May 13.  Received 2009 May 12; in original form 2009 January 29}
\author[J. Heyl and M. Penrice]{Jeremy Heyl$^{1\dagger}$ and Matt Penrice$^1$\\
$^{1}$Department of Physics and Astronomy, University of British Columbia, Vancouver, British Columbia, Canada, V6T 1Z1 \\
$^{\dagger}$Email: heyl@phas.ubc.ca; Canada Research Chair}

\pagerange{\pageref{firstpage}--\pageref{lastpage}} \pubyear{2009}

\maketitle

\label{firstpage}

\begin{abstract}
  Observations of white dwarfs in the globular clusters NGC 6397 and
  Omega Centauri indicate that these stars may get a velocity kick
  during their time as giants.  If the mass loss while on the
  asymptotic giant branch is slightly asymmetric, the resulting white
  dwarf could be born with such a velocity kick.  These energetic
  white dwarfs will impart their excess energy on other stars as they
  travel through the cluster.  A Monte-Carlo simulation of the
  white-dwarfs kicks combined with estimate of the phase-space
  diffusion of the white dwarfs reveals that as the white dwarfs
  equilibrate, they lose most of their energy in the central region of
  the cluster.  They could possibly augment the effect of binaries,
  delaying core collapse or increasing the size of globular cluster cores.
\end{abstract}

\section{Introduction}
\label{sec:introduction}

Several phenomena give circumstantial evidence that white dwarfs may
receive a velocity boost (or kick) at birth.  The observed rotation
rates of white dwarfs could result from mild kicks ($\sim$ km/s)
generated by asymmetric and off-centered winds toward the end of their
time on the asymptotic giant branch (AGB) \citep{1998A&A...333..603S}.
It is unclear at present how white dwarfs could actually obtain such a
kick; however, given the evidence it is natural to explore the
consequences of white-dwarf kicks.  Mild kicks may explain the
possible lack of white dwarfs in open clusters
\citep{2003ApJ...595L..53F,1977A&A....59..411W,2001AJ....122.3239K}.
Most directly, \citet{2008MNRAS.383L..20D} observed that the young
white dwarfs in NGC~6397 had a more expansive radial distribution than
either their progenitors or older white dwarfs.
\citet{2008MmSAI..79..347C} found similar but weaker hints in Omega
Centauri.

Without a kick young white dwarfs would have a velocity distribution
nearly equal to that of their more massive progenitors on the main
sequence.  In this case the kinetic energy of these white dwarfs is
much less than equipartition; therefore, as their velocity
distribution relaxes they cool the rest of the cluster.  If on the
other hand white dwarfs receive a substantial kick at birth as
observations may indicate \citep{2008MNRAS.383L..20D}, young white
dwarfs may heat the rest of the stars in the cluster.  Neutron stars
that form in the cluster may get a very large velocity kick as well;
however, two factors make their contribution less important.  The
first is that neutron stars are relatively rare, and the second is
that neutron stars travel so quickly that most escape the cluster
before imparting much energy to the rest of the stars
\citep{2002ApJ...573..283P}.  In contrast with the neutron-star result
\citet{Heyl07kickgc} found that most of the kicked white dwarfs remain
in the cluster.  \citet{Heyl08kickheat} found that the energy input
from the white-dwarf kicks around the half-light radius is about half
that from binaries at the present day and dominated in the first half
of the life of the cluster. This letter examines where in the cluster
white dwarfs dump their excess energy from the kicks.

\section{Calculations}
\label{sec:calculations}

A globular cluster is often modelled with a lowered isothermal
profile (or King model) \citep{1963MNRAS.125..127M,1966AJ.....71...64K,Binn87},
\begin{equation}
f = \frac{\d N}{\d^3 x \d^3 p} = 
 \left \{ 
\begin{array}{ll}
\rho_1(2\pi\sigma^2)^{-3/2} \left ( e^{\epsilon/\sigma^2}- 1 \right )
&  \rmmat{~if~} \epsilon>0 \\
0 & \rmmat{~if~} \epsilon\leq 0 
\end{array}
\right .
\label{eq:1}
\end{equation}
where $\epsilon = \Psi - \frac{1}{2} v^2$ and $\Psi$ is the
gravitational potential, $\sigma$ is the velocity dispersion of the
cluster stars, and $\rho_1$ is a characteristic number density.  The
distribution function depends only on the energy, $-m \epsilon$, a
constant of the motion; therefore, it is constant in time as well.

As the stars interact with each other, their kinetic energy approaches
equipartition such that $m_i \sigma_i^2 = m_j \sigma_j^2$ for masses
$m_i$ not equal to $m_j$ \citep{Spit87}.  The most massive
main-sequence stars in the cluster are the progenitors of the white
dwarfs so so they will typically have
$\sigma_\rmscr{TO}<\sigma_\rmscr{cluster}$, where
$\sigma_\rmscr{cluster}$ is the mean velocity dispersion of the
cluster. To model the cluster, the turn-off stars and the white-dwarf
kicks, we use the best-fitting model from \citet{Heyl07kickobs}.
Specifically, we take $\sigma_\rmscr{cluster}=1$ for the cluster
stars, $\sigma_\rmscr{TO}=0.5$ for the turn-off stars, the central
potential to $\Psi(0)=6$ and the mass of the white dwarfs to be 1.5
times the mass of the typical cluster member.  We model the kick as a
Gaussian in velocity with a dispersion of $\sigma_k=0.92$.  This means
that the kick is similar to the velocity dispersion of the cluster.
Although this is reasonable for clusters such as NGC 6397 with
$\sigma_\rmscr{cluster}\approx 5$~km/s, it is more difficult to
achieve for clusters such as 47~Tuc and Omega~Cen with
$\sigma_\rmscr{cluster} > 20$~km/s.

\subsection{Relaxation from a kick}

The rate that a particular star gains energy as it passes through the
cluster is given by \citep{Binn87}
\begin{eqnarray}
  D(\Delta E) &=& 16 \pi^2 G^2 m m_a^2 \ln \Lambda \times \nonumber \\
  & &  ~~\left [ \int_v^\infty
    v_a f_a(v_a) dv_a - \frac{m}{m_a}  \int_0^v \frac{v_a^2}{v} f_a(v_a) dv_a \right
  ].
\end{eqnarray}
where $m$ and $v$ are the mass and velocity of the star, $m_a$ and
$v_a$ are the mass and velocity of the other cluster members,
$\ln\Lambda$ is the Coulomb logarithm and
$f_a(v_a)$ is the phase-space density. The first term is the energy
that a particular star gains from encounters from faster moving stars.
For a King model it is
\begin{equation}
\int_v^\infty v_a f_a(v_a)dv_a = \rho_1 \left(2\pi\sigma^2\right)^{-3/2}
  \left [ \sigma^2 \left ( e^{\epsilon/\sigma^2} - 1  \right ) -
    \epsilon \right ]
\end{equation}
where $\epsilon = \Psi - \frac{1}{2} v^2$ for the particular star,
$\sigma$ is the dispersion for the cluster as a whole.
The second term is the energy that the star loses from encounters with
slower moving stars,
\begin{eqnarray}
\frac{m}{m_a} \int_0^v \frac{v_a^2}{v} f_a(v_a) dv_a &=& 
\rho_1 \frac{m}{m_a} \left(2\pi\sigma^2\right)^{-3/2} \times \\
& &  
\!\!\!\!\!\!\!\!\!\!\!\!\!\!\!\!\!\!\!\!\!\!\!\!\!\!\!\!\!\!\!\!\!\!\!
\left [ \sqrt{\frac{\pi}{2}} \frac{\sigma^3}{v} e^{\Psi/\sigma^2} {\rm erf} \left (
  \frac{v}{\sqrt{2} \sigma} \right ) - 
\left (e^{\epsilon/\sigma^2}\sigma^2 + \frac{v^2}{3}\right ) \right ] .\nonumber
\end{eqnarray}
Because we know the position and velocity of each of the white dwarfs
in the simulation, it is straightforward to calculate the energy loss
and gain from these equations.

\section{Results}
\label{sec:results}

\begin{figure}
 \includegraphics[width=3.4in]{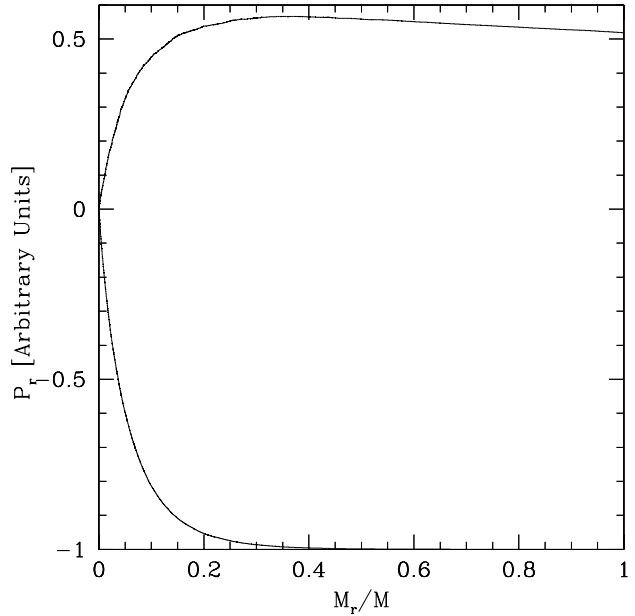}
 \caption{The power imparted to the cluster from young white dwarfs
   within a given radius as a function of the enclosed cluster mass.
   The upper curve traces the value for white dwarfs with a kick.}
 \label{fig:big}
\end{figure}

To explore the distribution of the energy from the kicks to the rest of
the cluster members, we created a Monte-Carlo realisation of 100,000
newly born white dwarfs both with and without kicks \citep[see][for
further details]{Heyl07kickgc}.  For each realisation the white dwarfs
are sorted by radius and the total energy deposited into the cluster
stars is summed by radius resulting in the cumulative distribution of
power depicted in Fig.~\ref{fig:big}.  The upper curve traces the
result for white dwarfs with kicks.  The white dwarfs are travelling
too fast for their mass so they deposit the excess energy into the
cluster.  The lower curve shows the result for white dwarfs without a
kick --- here, the white dwarfs are born with the velocity dispersion
of their progenitors who are significantly more massive than
themselves; therefore, the white dwarfs sap energy from the cluster
stars as they heat up.  The exact locations of the two curves depend
on the assumed value of the ratio between the white-dwarf mass and
that of the cluster stars (the figure gives the result for 1.5; as the
ratio increases both curves move up).

\begin{figure}
 \includegraphics[width=3.4in]{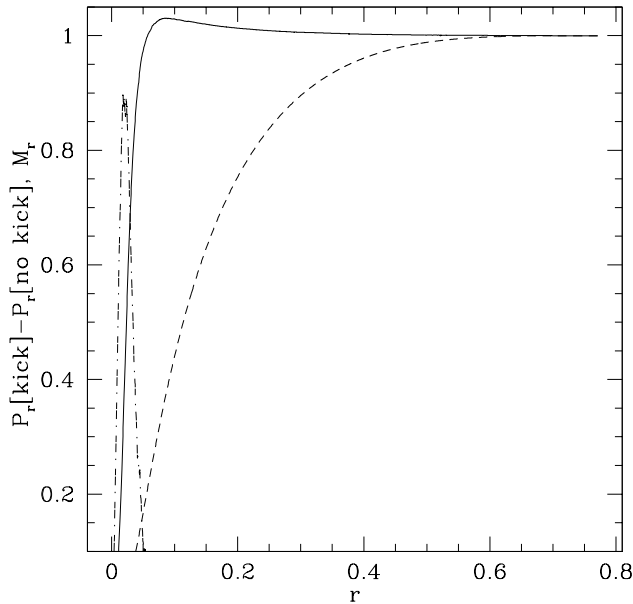}
 \caption{Net increase in energy deposition from the kicks within a
   given radius as a function of radius and the mass enclosed with
   that radius.  The dashed curve panel traces the cumulative mass
   distribution.  Both the cumulative (solid curve) and the
   differential power distributions (dot-dashed curve) are shown. The
   tidal radius of the cluster is $r_t=0.77$, the core radius is
   $r_c=0.04$, and the half-light radius is $r_h=0.11$.  Ninety
   percent of the energy is deposited within a core radius of the
   centre of the cluster.  Beyond a radius of 0.14, the kicked white
   dwarfs begin to lose energy to the cluster stars.}
 \label{fig:diff}
\end{figure}

The distance between the cumulative energy deposition rates is the net
effect of the white-dwarf kicks.   This net effect is rather robust
because regardless of the assumed ratio between the white dwarf masses
and the rest of the stars, white dwarfs are typically created near
the centre of the cluster because their progenitors are massive.
Furthermore, even with a kick, the white dwarfs will return to the
radius of their birth until their orbits relax.   At this small
radius, the speed of the white dwarf is largest and so is the density
of the rest of the stars; therefore, at small radii the white dwarfs
lose most of their excess energy.   Fig.~\ref{fig:diff}
demonstrates this fact.   For the assumed model about ninety percent
of the energy of the white-dwarf kicks is deposited within the core
radius, and the white dwarfs actually remove energy from the outer
regions of the cluster.

During the collapse of a globular cluster core, two-body interactions
result in the transfer of energy from the dense central region of the
cluster to the outer regions; the core shrinks, and the envelope of
the cluster expands.   The white dwarfs deposit their energy in
precisely the opposite way, supplying energy to the core at the
expense of the outer regions; therefore, white-dwarf kicks may retard
core collapse or increase the size of cluster cores.

\section{Conclusions}
\label{sec:conclusions}
 
After a globular cluster forms, the negative heat capacity of the
cluster's self-gravity inexorably draws the cluster toward core
collapse, unless another energy source is present.  The central region
or core will collapse to very large densities within a few relaxation
times while the outer regions expand, and stars evaporate from the
system.  The standard energy source that is invoked to retard the
collapse are binaries \citep[e.g][]{1989Natur.339...40G}.  After the
core has already reached high densities, binaries necessarily form
through tidal and three-body interactions preventing the core from
forming a singularity\citep[e.g][]{2007ApJ...665..707H}.  

$N$-body models with a primordial binary fraction of a few percent can
avoid the complete collapse of the core but the core does contract
dramatically over the age of the cluster.  If all old globular
clusters had compact cores, this scenario would not
pose a problem; however, several old, presumably relaxed clusters have
large cores without an obvious energy source to explain the observed state
\citep{2007ApJ...656L..65D}.  Several possibilities have been invoked
to explain this: a very high initial binary fraction approaching unity
(however, \citet{2006MNRAS.368..677H} found that the effect of
binaries saturates at about ten percent), an intermediate-mass black
hole \citep[e.g.][]{2007MNRAS.374..857T} or a black-hole binary
\citep{2008MNRAS.386...65M} at the centre of the cluster.  As for the
white-dwarf kicks, the evidence for these possibilities is
circumstantial.

There are several lines of evidence that white dwarfs get a mild
velocity kick at birth.  Furthermore, the total power deposited into
the cluster by the white-dwarf kicks may be comparable or larger than
binaries (\citealt{Heyl08kickheat}: although these calculations should be
generalised to the core for completeness) and the energy is deposited
in such a way that it directly opposes core collapse, so these kicks
may provide a explanation for the fraction of globular cluster that
have large cores today.  We do however see a variety of core sizes
including collapsed cores, so the combination of white-dwarf kicks and
binaries must not always be effective or perhaps the clusters are at
various stages of their evolution.  Explaining these differences in the
white-dwarf kick model is difficult as one would expect that all
globular clusters should exhibit the effects of the kicks.  On the
other hand the black hole models can easily explain these differences;
some clusters could have a large black hole at their centres and
others lack one.

The Monte-Carlo model present here is illustrative, but rather
rudimentary.  Two obvious avenues for further study are the inclusion
of white-dwarf kicks in $N$-body models
\citep[e.g.][]{2007ApJ...665..707H} and Monte Carlo simulations
\citep[e.g.][]{2008MNRAS.389.1858H} --- either of these avenues would
allow us to include both binaries and white dwarf kicks together
consistently to get a more robust estimate of their relative
importance.  Furthermore, one could use a more realistic multi-mass
King model for the background stars in the cluster to perform an
analysis similar to that presented here.  One would expect that that
results could differ quantitatively but the qualitative result the
white-dwarfs kicks dump energy in the central regions of the cluster
and that this energy may be comparable or larger than that from
binaries is robust.

\section*{Acknowledgments}

J.H. like to thank Harvey Richer for useful discussions.  We would
also like to thank the reviewer for many constructive comments. The
Natural Sciences and Engineering Research Council of Canada, Canadian
Foundation for Innovation and the British Columbia Knowledge
Development Fund supported this work.  Correspondence and requests for
materials should be addressed to heyl@phas.ubc.ca.  This research has
made use of NASA's Astrophysics Data System Bibliographic Services

\bibliographystyle{mn2e}
\bibliography{mine,wd,physics,math}
\label{lastpage}
\end{document}